\documentstyle[axodraw,12pt,a4]{article}
\newcommand\inp{{\cdot}}
\newcommand\GeV{\mbox{ GeV}}
\newcommand\GEV{\mbox{GeV}}
\newcommand\pB{\mbox{ pb}}
\newcommand\PB{\mbox{pb}}
\newcommand\mg{m_{\gamma\gamma}}

\newcommand{\be}{\begin{equation}}
\newcommand{\ee}{\end{equation}}
\newcommand{\ba}{\begin{eqnarray}}
\newcommand{\ea}{\end{eqnarray}}
\newcommand{\baa}{\begin{eqnarray*}}
\newcommand{\eaa}{\end{eqnarray*}}
\newcommand{\bb}{\begin{thebibliography}}
\newcommand{\eb}{\end{thebibliography}}
\newcommand{\ci}[1]{\cite{#1}}
\newcommand{\bi}[1]{\bibitem{#1}}

\newcommand{\bit}{\begin{itemize}}
\newcommand{\eit}{\end{itemize}}
\newcommand{\g}{\gamma}

\newcommand{\epmmgg}{e^+e^- \rightarrow \mu^+\mu^- \gamma \gamma}
\newcommand{\epllgg}{e^+e^- \rightarrow \ell^+\ell^- \gamma \gamma}
\newcommand{\gv}{\mbox{GeV}}
\def\setepsfscale#1{\def\epsfsize##1##2{#1##1}}
\begin{document}
\thispagestyle{empty}
\hfill {\sc BI-TP-93/01} \qquad { } \par
\hfill {\sc PSI-PR-93-01} \qquad { } \par
\hfill January, 1993 \qquad { }

\vspace*{15mm}

\begin{center}
{\LARGE Production of two hard isolated photons in $\epmmgg$ at LEP}

\medskip

\vspace{16mm}

{\large   K. ~Ko\l odziej \footnote
{On leave from the Institute of Physics, University of Silesia,
 40-007 Katowice, ul. Uniwersytecka 4, Poland}}

\medskip
{\em   Fakult\"at f\"ur Physik, Universit\"at Bielefeld
                  D-4800 Bielefeld 1, Germany}

\bigskip
{\large F.~Jegerlehner and G. J. van ~Oldenborgh}

\medskip

{\em Paul Scherrer Institute,  CH-5232 Villigen PSI, Switzerland}
\bigskip

\end{center}

\vspace*{20mm}

\textwidth 120mm
\centerline{Abstract}
\noindent
\normalsize
\setlength{\baselineskip}{0.85cm} 
A calculation of the hard bremsstrahlung process $\epmmgg$
is presented and compared with recent results from the L3 Collaboration.

\vfill

\newpage


\section{Introduction}
\setlength{\baselineskip}{0.85cm} 

\medskip

Recently the L3 collaboration \ci{[1]} has reported the observation of four
$\epllgg$ events where besides the lepton pair two hard photons at large angles
are produced. A clustering in the two-photon invariant mass $M_{\g \g}$ around
60 \gv \ was observed which seems unlikely to be explained by QED. In order
to establish possible new physics we first need a precise understanding of
the QED background. For this purpose we present a Monte Carlo event generator
which allows to take into account the appropriate detector efficiencies and
phase-space cuts.  We do not include the t-channel graphs which are needed for
the $e^+e^-\gamma\gamma$ final state.  Also, though mass effects are included,
we do not study $\tau$'s as these require a detailed simulation of the
detector.

Since all events have a reconstructed mass of the final state very close to
the $Z$ mass the main QED background must be due to double final state
bremsstrahlung from the produced lepton-pair. As the cross section varies
strongly near the $Z$ peak we take into account initial state bremsstrahlung
and initial-final state interference as well. The double hard photon
matrix element is convoluted with the full $O(\alpha^2)$ leading logarithms
improved single hard photon spectrum which describes the $Z$ resonance very
well \ci{[2]}.

Similar calculations were performed by other authors \ci{[3],[4]},
where ref. \ci{[4]} takes into account final state radiation only
which should provide a reasonably good approximation.  In the L3 article the
QED prediction was generated using the Yennie-Frautschi-Suura scheme
Monte Carlo YFS3
\cite{[5]}; the approximations made here may not be valid for the very
acollinear photons observed.  Because the issue
seems to be somewhat controversial we recalculated the process and we
present our results in the following.

We have calculated the matrix element for the process $\epmmgg$ using the
general method which was described in Ref. \ci{[6]}. A comparison of the
hard bremsstrahlung calculations for other processes showed that the
numerical sandwiching of $\gamma$-operators between helicity spinors is
much faster in computer time than the numerical evaluation of analytic
expressions obtained by means of the Weyl-van der Waerden formalism \ci{[7]}.
This note is organized as follows. In sect. 2 we describe the calculation of
the matrix element. The kinematics and the event generation is discussed in
sect. 3 . Results and conclusions follow in sects. 4 and 5, respectively.

\section{The Matrix Element}
\label{sec:ME}

In this section we will present briefly a method of calculating
a helicity matrix element of the process
\begin{equation}
\label{eq:process}
 e^+(p_+) + e^-(p_-) \rightarrow \mu^+(q_+) + \mu^-(q_-)
   + \gamma_1(k_1) + \gamma_2(k_2),
\end{equation}
where in parenthesis we indicate particle 4-momenta. The method
used here is an extension of the method originally developed in
\ci{[6]}
for the purely bososnic final states in $e^+e^-$ annihilation.
For the moment we assume that both initial and final
fermions are massless. This is justified
since we are interested in the kinematical configuration where
the photons are radiated at large angles with respect to the
initial electrons and final muons. The assumption makes our
method much simpler and speeds up computation of the matrix
element.  Later we will correct for the largest mass effects.

For massless fermions it is natural to use the Weyl
representation in which the Dirac $\gamma$ matrices read
\begin{equation}
\label{eq:rep}
 \gamma_\mu = \left(
 \begin{array}{ll}
           0  & \sigma_\mu\\
       {\tilde{\sigma}}_\mu & 0
 \end{array}
                \right),\qquad
 \gamma_{5} = \left(
 \begin{array}{cc}
          -1  & 0 \\
           0  & 1
 \end{array}
                \right),
\end{equation}
where, in terms of the Pauli matrices,
$ \sigma_\mu = \left(1,\vec{\sigma} \right)$ and
$ {\tilde{\sigma}}_\mu = \left(1,- \vec{\sigma} \right)$.
In this representation the helicity spinors for a particle
$u(\vec{p},\sigma)$ and an antiparticle $v(\vec{p},\sigma)$
(in the massless limit) are
\begin{eqnarray}
\label{eq:spin1}
 u(\vec{p},\sigma) & = & \sqrt{E} \left(
 \begin{array}{c}
           \sqrt{1 - \sigma}\quad \chi(\vec{p},\sigma) \\
           \sqrt{1 + \sigma}\quad \chi(\vec{p},\sigma)
 \end{array}
                \right),\\
\label{eq:spin2}
 v(\vec{p},\sigma) & = & \sigma \sqrt{E} \left(
 \begin{array}{c}
         - \sqrt{1 + \sigma}\quad \chi(\vec{p},- \sigma) \\
           \sqrt{1 - \sigma}\quad \chi(\vec{p},- \sigma)
 \end{array}
                \right),
\end{eqnarray}
where the Pauli helicity spinors $\chi(\vec{p},\sigma)$ for
either a particle or an antiparticle with momentum
$\vec{p} = (E,\theta,\phi)$ and the helicity $\sigma / 2 =
\pm 1/2$ are
\begin{equation}
\label{eq:pauli}
 \chi (\vec{p},+1) = \left(
 \begin{array}{c}
     e^{-i\phi /2}\cos \theta / 2 \\
     e^{i\phi /2}\sin \theta / 2
 \end{array}
                \right),\qquad
 \chi (\vec{p},-1) = \left(
 \begin{array}{c}
   - e^{-i\phi /2}\sin \theta / 2 \\
     e^{i\phi /2}\cos \theta / 2
 \end{array}
                \right).
\end{equation}
We use the spinors of eq.\ (\ref{eq:pauli}) also for particles going into
$-\vec{p}$ direction which means that we do not stick to the
Jacob and Wick convention \cite{Jacob&Wick}.

For the sake of simplicity and higher numerical speed
we take photon polarisation vectors in the rectangular
basis where for the photon with momentum $\vec{k} =
(k,\theta ,\phi)$
we have
\begin{equation}
 \epsilon^{\mu}(\vec{k},1) = \left(
 \begin{array}{c}
                  0 \\
                 \cos \theta \cos \phi \\
                 \cos \theta \sin \phi \\
                -\sin \theta
 \end{array}
                \right),\qquad
 \epsilon^{\mu}(\vec{k},2) = \left(
 \begin{array}{c}
                  0 \\
                -\sin \phi \\
                 \cos \phi \\
                  0
 \end{array}
                \right).
\end{equation}

In the representation of eq.\ (\ref{eq:rep}) for any 4-vector $a^{\mu}$ we have
\begin{equation}
\label{eq:ahat}
 \hat{a} = a^{\mu}\gamma_\mu = \left(
 \begin{array}{cc}
                  0 & a^{\mu}\sigma_\mu\\
          a^{\mu}{\tilde{\sigma}}_\mu & 0
 \end{array}
                \right),
\end{equation}
where in terms of components (real or complex)
\begin{eqnarray}
 a & \equiv & a^{\mu}\sigma_\mu = \left(
 \begin{array}{cc}
          a^0 - a^3 & -a^1 + ia^2\\
          -a^1 - ia^2 & a^0 + a^3
 \end{array}
                \right), \\
 \tilde{a} & \equiv & a^{\mu}{\tilde{\sigma}}_\mu = \left(
 \begin{array}{cc}
          a^0 + a^3 & a^1 - ia^2\\
          a^1 + ia^2 & a^0 - a^3
 \end{array}
                \right).
\end{eqnarray}
Thus the product of any odd number of matrices of eq.
(\ref{eq:ahat}) has antidiagonal form
\begin{equation}
\label{eq:abc}
 \hat{a}\hat{b}\hat{c}\dots = \left(
 \begin{array}{cc}
                  0 & a\tilde{b}c\dots \\
          \tilde{a}b\tilde{c}\dots & 0
 \end{array}
                \right) \equiv \left(
 \begin{array}{cc}
                  0 & A \\
              \tilde{A} & 0
 \end{array}
                \right) \equiv \left(\hat{A} \right)\;\;\;.
\end{equation}

Our method of calculation is simple because of the fact that,
in the massless case,
strings of Dirac matrices of eq.\ (\ref{eq:abc}) are sandwiched between
spinors of eqs (\ref{eq:spin1}) and (\ref{eq:spin2}) which, for a given
helicity, are half-empty. This reduces $4 \times 4$ to $2 \times 2$
matrix algebra.  Moreover, if the particles are moving along the
$z$-axis there is only one nonvanishing element in each of the spinors
of eq.\ (\ref{eq:pauli}).

\begin{figure}[htb]
\def\axoscale{.7 }
\unitlength .7bp
\centerline{
\begin{picture}(100,100)(0,0)
\ArrowLine(30,30)(0,0)
\ArrowLine(0,90)(30,30)
\Photon(30,30)(70,30){4}{5}
\Vertex(30,30){1.0}
\Vertex(70,30){1.0}
\ArrowLine(100,0)(70,30)
\ArrowLine(70,30)(100,90)
\Photon(10,70)(40,100){4}{3}
\Vertex(10,70){1.0}
\Photon(20,50)(50,80){4}{3}
\Vertex(20,50){1.0}
\put(0,30){\makebox(0,0)[b]{I1}}
\put(0,0){\makebox(0,0)[br]{$e^+$}}
\put(0,90){\makebox(0,0)[tr]{$e^-$}}
\put(50,24){\makebox(0,0)[t]{$\gamma,Z$}}
\put(40,100){\makebox(0,0)[bl]{$\gamma$}}
\put(50,80){\makebox(0,0)[bl]{$\gamma$}}
\put(100,0){\makebox(0,0)[bl]{$\mu^+$}}
\put(100,90){\makebox(0,0)[tl]{$\mu^-$}}
\end{picture}
\hspace{1.3cm}
\begin{picture}(100,100)(0,0)
\ArrowLine(30,45)(15,22.5)
\ArrowLine(15,22.5)(0,0)
\ArrowLine(0,90)(15,67.5)
\ArrowLine(15,67.5)(30,45)
\Photon(30,45)(70,45){4}{5}
\Vertex(30,45){1.0}
\Vertex(70,45){1.0}
\ArrowLine(100,0)(70,45)
\ArrowLine(70,45)(100,90)
\Photon(15,67.5)(45,97.5){4}{3}
\Vertex(15,67.5){1.0}
\Photon(15,22.5)(45,-7.5){4}{3}
\Vertex(15,22.5){1.0}
\put(0,45){\makebox(0,0){I2}}
\end{picture}
\hspace{1.3cm}
\begin{picture}(100,100)(0,0)
\ArrowLine(30,60)(0,0)
\ArrowLine(0,90)(30,60)
\Photon(30,60)(70,60){4}{5}
\Vertex(30,60){1.0}
\Vertex(70,60){1.0}
\ArrowLine(100,0)(70,60)
\ArrowLine(70,60)(100,90)
\Photon(10,20)(40,-10){4}{3}
\Vertex(10,20){1.0}
\Photon(20,40)(50,10){4}{3}
\Vertex(20,40){1.0}
\put(0,60){\makebox(0,0)[t]{I3}}
\end{picture}
}
\bigskip
\centerline{
\begin{picture}(100,100)(0,0)
\ArrowLine(30,45)(0,0)
\ArrowLine(0,90)(15,67.5)
\ArrowLine(15,67.5)(30,45)
\Photon(30,45)(70,45){4}{5}
\Vertex(30,45){1.0}
\Vertex(70,45){1.0}
\ArrowLine(100,0)(85,22.5)
\ArrowLine(85,22.5)(70,45)
\ArrowLine(70,45)(100,90)
\Photon(15,67.5)(45,97.5){4}{3}
\Vertex(15,67.5){1.0}
\Photon(85,67.5)(115,50){4}{3}
\Vertex(85,67.5){1.0}
\put(0,45){\makebox(0,0){IF1}}
\end{picture}
\hspace{1.1cm}
\begin{picture}(100,100)(0,0)
\ArrowLine(30,45)(0,0)
\ArrowLine(0,90)(15,67.5)
\ArrowLine(15,67.5)(30,45)
\Photon(30,45)(70,45){4}{5}
\Vertex(30,45){1.0}
\Vertex(70,45){1.0}
\ArrowLine(100,0)(85,22.5)
\ArrowLine(85,22.5)(70,45)
\ArrowLine(70,45)(100,90)
\Photon(15,67.5)(45,97.5){4}{3}
\Vertex(15,67.5){1.0}
\Photon(85,22.5)(115,40){4}{3}
\Vertex(85,22.5){1.0}
\put(0,45){\makebox(0,0){IF2}}
\end{picture}
\hspace{1.1cm}
\begin{picture}(100,100)(0,0)
\ArrowLine(30,45)(15,22.5)
\ArrowLine(15,22.5)(0,0)
\ArrowLine(0,90)(30,45)
\Photon(30,45)(70,45){4}{5}
\Vertex(30,45){1.0}
\Vertex(70,45){1.0}
\ArrowLine(100,0)(85,22.5)
\ArrowLine(85,22.5)(70,45)
\ArrowLine(70,45)(100,90)
\Photon(15,22.5)(45,-7.5){4}{3}
\Vertex(15,22.5){1.0}
\Photon(85,67.5)(115,50){4}{3}
\Vertex(85,67.5){1.0}
\put(0,45){\makebox(0,0){IF3}}
\end{picture}
\hspace{1.1cm}
\begin{picture}(100,100)(0,0)
\ArrowLine(30,45)(15,22.5)
\ArrowLine(15,22.5)(0,0)
\ArrowLine(0,90)(30,45)
\Photon(30,45)(70,45){4}{5}
\Vertex(30,45){1.0}
\Vertex(70,45){1.0}
\ArrowLine(100,0)(85,22.5)
\ArrowLine(85,22.5)(70,45)
\ArrowLine(70,45)(100,90)
\Photon(15,22.5)(45,-7.5){4}{3}
\Vertex(15,22.5){1.0}
\Photon(85,22.5)(115,40){4}{3}
\Vertex(85,22.5){1.0}
\put(0,45){\makebox(0,0){IF4}}
\end{picture}
}
\bigskip
\centerline{
\begin{picture}(100,100)(0,0)
\ArrowLine(30,30)(0,0)
\ArrowLine(0,90)(30,30)
\Photon(30,30)(70,30){4}{5}
\Vertex(30,30){1.0}
\Vertex(70,30){1.0}
\ArrowLine(100,0)(70,30)
\ArrowLine(70,30)(100,90)
\Photon(90,70)(120,40){4}{3}
\Vertex(90,70){1.0}
\Photon(80,50)(110,20){4}{3}
\Vertex(80,50){1.0}
\put(0,30){\makebox(0,0)[b]{F1}}
\end{picture}
\hspace{1.3cm}
\begin{picture}(100,100)(0,0)
\ArrowLine(30,45)(0,0)
\ArrowLine(0,90)(30,45)
\Photon(30,45)(70,45){4}{5}
\Vertex(30,45){1.0}
\Vertex(70,45){1.0}
\ArrowLine(100,0)(85,22.5)
\ArrowLine(85,22.5)(70,45)
\ArrowLine(70,45)(85,67.5)
\ArrowLine(85,67.5)(100,90)
\Photon(85,67.5)(115,50){4}{3}
\Vertex(85,67.5){1.0}
\Photon(85,22.5)(115,40){4}{3}
\Vertex(85,22.5){1.0}
\put(0,45){\makebox(0,0){F2}}
\end{picture}
\hspace{1.3cm}
\begin{picture}(100,100)(0,0)
\ArrowLine(30,60)(0,0)
\ArrowLine(0,90)(30,60)
\Photon(30,60)(70,60){4}{5}
\Vertex(30,60){1.0}
\Vertex(70,60){1.0}
\ArrowLine(100,0)(70,60)
\ArrowLine(70,60)(100,90)
\Photon(90,20)(120,50){4}{3}
\Vertex(90,20){1.0}
\Photon(80,40)(110,70){4}{3}
\Vertex(80,40){1.0}
\put(0,60){\makebox(0,0)[t]{F3}}
\end{picture}
}
\unitlength 1bp
\def\axoscale{1. }
\caption{Half the Feynman diagrams for the reaction
$e^+e^-\to\mu^+\mu^-\gamma\gamma$, the other half
follow by crossing the final state photons.}
\label{fig:feynman}
\end{figure}

Feynman diagrams for the process under consideration are shown in fig.\
\ref{fig:feynman}. Together with the corresponding crossed diagrams
they form three gauge independent subsets which correspond
to the radiation of photons from the initial state \{I\}, final state \{F\}
and from the initial and final state \{IF\}. We will explain the method by
following the calculation of the matrix element corresponding to diagram I1 in
fig.\ \ref{fig:feynman}.
\begin{equation}
M_{I1}(\sigma,\bar{\sigma},\lambda_1,\lambda_2) =
M_{I1}^{\gamma}(\sigma,\bar{\sigma},\lambda_1,\lambda_2) +
M_{I1}^Z(\sigma,\bar{\sigma},\lambda_1,\lambda_2),
\end{equation}
where $\sigma / 2$ ($\bar{\sigma} / 2$) is the helicity of the electron
(of the $\mu^-$) and $\lambda_{1,2}$ are the polarisations of the photons.
Since the fermions are massless nonvanishing matrix elements
are obtained only if the helicity of the positron (of the $\mu^+$)
is opposite to the one of the electron (of the $\mu^-$). Hence,
\begin{eqnarray}
\label{eq:MI1}
\lefteqn{M_{I1}(\sigma,\bar{\sigma},\lambda_1,\lambda_2) \equiv
   C_I(\sigma,\bar{\sigma})
\bar{v}({\vec{p}}_+,-\sigma)\gamma^{\mu} ({\hat{p}}_- - {\hat{k}}_1
- {\hat{k}}_2) {\hat{\epsilon}}({\vec{k}}_2,\lambda_2)}\nonumber\\
&&\mbox{} \times ({\hat{p}}_- - {\hat{k}}_1)
{\hat{\epsilon}}({\vec{k}}_1,\lambda_1) P_{\sigma} u({\vec{p}}_-,\sigma)
\bar{u}({\vec{q}}_-,\bar{\sigma}) \gamma_\mu P_{\bar{\sigma}}
v({\vec{q}}_+,-\bar{\sigma})\nonumber\\
&\equiv & C_I(\sigma,\bar{\sigma})
\bar{v}({\vec{p}}_+,-\sigma)({\hat{A}}^{\mu}) P_{\sigma} u({\vec{p}}_-,\sigma)
\bar{u}({\vec{q}}_-,\bar{\sigma}) ({\hat{B}}_\mu) P_{\bar{\sigma}}
v({\vec{q}}_+,-\bar{\sigma}),
\end{eqnarray}
where $({\hat{A}}^{\mu})$ and $({\hat{B}}^{\mu})$ are antidiagonal matrices
of the form of eq.\ (\ref{eq:abc}), which correspond to a string of an odd
number of Dirac matrices with one uncontracted Lorentz index
sandwiched between the initial and final state spinors, respectively.
$P_{\sigma} \equiv { \frac{1}{2} (1 - \sigma \gamma_5)}$ are helicity
projectors. The amplitudes $C_I$ are defined by
\begin{eqnarray}
   C_I(\sigma,\bar{\sigma}) & \equiv &
    \frac{e^{4}}{(p_- - k_1)^2(p_- - k_1 - k_2)^2}\nonumber\\
&\times &\left[\frac{1}{{(q_+ + q_-)}^2} + \frac{c_I(\sigma,\bar{\sigma})}
{\left( {(q_+ + q_-)}^2 - M_Z^2 + iM_Z \Gamma_Z \right)} \right],
\end{eqnarray}
where
\begin{equation}
c_I(\sigma,\bar{\sigma}) \equiv \left\{
\begin{array}{cl}
     (g_V + g_A)^2  & \quad {\rm if}\quad \sigma = \bar{\sigma} = -1\\
     (g_V - g_A)^2  & \quad {\rm if}\quad \sigma = \bar{\sigma} = +1\\
     g_V^2 - g_A^2  & \quad {\rm if}\quad \sigma = -\bar{\sigma}\;\;\;\;\;\;.
\end{array} \right.
\end{equation}

\begin{eqnarray}
g_V = - {\frac{1-4\sin^2 \theta_W}{4 \sin \theta_W \cos \theta_W}}
\qquad {\rm and} \qquad
g_A = - {\frac{1}{4 \sin \theta_W \cos \theta_W}}
\end{eqnarray}
are the vector and axial-vector parts of the $eeZ$-coupling, respectively.

Now we observe that the only nonvanishing `sandwiches' in eq.\
(\ref{eq:MI1}) are (in obvious notation)
\begin{eqnarray}
\bar{v}({\vec{p}}_+,+)({\hat{A}}^{\mu}) P_- u({\vec{p}}_-,-) & = &
i2E \chi^{\dag}(\vec{p}_+,+)
{\tilde{A}}^{\mu} \chi (\vec{p}_-,-) = i2E {\tilde{A}}^{\mu}_{1,2},
\nonumber\\
\bar{v}(\vec{p}_+,-)({\hat{A}}^{\mu}) P_+ u(\vec{p}_-,+) & = &
i2E \chi^{\dag}(\vec{p}_+,-)
A^{\mu} \chi (\vec{p}_-,+) = i2E A^{\mu}_{2,1},
\end{eqnarray}
for the initial state fermions and
\begin{eqnarray}
\lefteqn{\bar{u}(\vec{q}_-,+)({\hat{B}}^{\mu}) P_+ v(\vec{q}_+,-)}\nonumber\\
 & = & -2\sqrt{E_-E_+}
\chi^{\dag}(\vec{q}_-,+) B^{\mu} \chi (\vec{q}_+,+) \nonumber\\
& = & -2\sqrt{E_-E_+} \left[
e^{-i{\frac{\phi_+ - \phi_-}{2}}} \cos {\frac{\theta_+}{2}}
\cos {\frac{\theta_-}{2}} B^{\mu}_{1,1}
+e^{i{\frac{\phi_+ + \phi_-}{2}}} \sin {\frac{\theta_+}{2}}
\cos {\frac{\theta_-}{2}} B^{\mu}_{1,2}
\right. \nonumber\\ &&
\left.
\mbox{} +e^{-i{\frac{\phi_+ + \phi_-}{2}}} \cos {\frac{\theta_+}{2}}
\sin {\frac{\theta_-}{2}} B^{\mu}_{2,1}
+e^{i{\frac{\phi_+ - \phi_-}{2}}} \sin {\frac{\theta_+}{2}}
\sin {\frac{\theta_-}{2}} B^{\mu}_{2,2}
\right], \nonumber\\
\lefteqn{\bar{u}(\vec{q}_-,-)({\hat{B}}^{\mu}) P_- v(\vec{q}_+,+)} \nonumber\\
 & = & -2\sqrt{E_-E_+}
\chi^{\dag}(\vec{q}_-,-) {\tilde{B}}^{\mu} \chi (\vec{q}_+,-) \nonumber\\
& = & -2\sqrt{E_-E_+} \left[
e^{-i{\frac{\phi_+ - \phi_-}{2}}} \sin {\frac{\theta_+}{2}}
\sin {\frac{\theta_-}{2}} {\tilde{B}}^{\mu}_{1,1}
-e^{i{\frac{\phi_+ + \phi_-}{2}}} \cos {\frac{\theta_+}{2}}
\sin {\frac{\theta_-}{2}} {\tilde{B}}^{\mu}_{1,2}
\right. \nonumber\\
&& \left.
-e^{-i{\frac{\phi_+ + \phi_-}{2}}} \sin {\frac{\theta_+}{2}}
\cos {\frac{\theta_-}{2}} {\tilde{B}}^{\mu}_{2,1}
+e^{i{\frac{\phi_+ - \phi_-}{2}}} \cos {\frac{\theta_+}{2}}
\cos {\frac{\theta_-}{2}} {\tilde{B}}^{\mu}_{2,2}
\right]
\end{eqnarray}
for the final state fermions, respectively.
In order to obtain the polarised matrix element corresponding to
diagram I1 in fig.\ \ref{fig:feynman} it is sufficient to calculate one
element of the $2 \times 2$ matrix $A^{\mu}$ or
${\tilde{A}}^{\mu}$ for $\mu = 0,1,2,3$ and then, after contracting
the index $\mu$, the complete $2 \times 2$ matrix
$B$ or $\tilde{B}$. We repeat the same procedure for each
Feynman diagram in fig.\ \ref{fig:feynman}. Appropriate routines are
implemented in a numerical computer program which computes the matrix
element squared averaged over the spins.

To check our calculation we performed several tests. First of
all we tested electromagnetic gauge invariance with respect to each
of the photons for each of the gauge invariant subsets
of diagrams \{I\}, \{F\} and \{IF\}, separately. We replaced the
photon polarisation vector by its 4-momentum and observed numerically
a drop in the real and imaginary parts of the matrix element by
15 -- 20 orders of magnitude for each polarisation.
For the final state diagrams \{F\} without the $Z$ we compared the results
for the squared matrix element with the ones obtained using the
traditional trace
technique employing the algebraic manipulation program FORM \cite{Form}. Both
results agree up to 14 digits.

Finally, we can introduce the largest (logarithmic) mass effects by replacing
the massless collinearly divergent propagators in the final state with the
corresponding massive ones.  A gauge invariant way to do that is to multiply
the
entire matrix element squared by the ratio $q_\pm \inp k_i/\tilde{q}_\pm \inp
\tilde{k}_i$, where the momenta $\tilde{q}_\pm$, $\tilde{k_i}$ refer to a set
of four momenta constructed without neglecting the masses.  Another overall
factor is necessary to treat the low invariant mass muon pairs
correctly, here the relevant ratio is $q_+ \inp
q_-/(\tilde{q}_+\tilde{q}_- + m_\mu^2)$.


\section{Kinematics}
\label{sec:kin}

In this section we first describe how to construct the kinematics for massive
muons.  The corresponding momenta are used for the cuts and the phase space
factors.  Using the same values for the integration variables (adjusted
for the different boundaries) we also construct a set of massless
momenta which are used in the matrix element. Using the overall factors
described before this will approximate most mass effects.  From here on
we will leave out the tildes in the massive vectors.

The matrix element squared has single poles in $q_\pm \inp k_1$, $q_\pm
\inp k_2$, $p_\pm \inp k_1$ and $p_\pm \inp k_2$ for massless fermions%
\footnote{The double poles have a numerator proportional to the mass of the
fermions and can be neglected for the cuts in \cite{[1]}.}.  With
Monte Carlo integration routine VEGAS \cite{[10]} we will have to make these
variables proportional to the integration variables and map them away to get a
reasonable accuracy.  However, this is not possible as there are more poles
(8) than integration variables (7, not counting the rotation around the beam
axis $\phi$, on which the matrix element does not depend for unpolarized
beams).  Since the angle
between the photons and the beam direction is required to be quite large in
\ci{[1]} ($|\cos\theta_{e\gamma}|<.9$), we can neglect the effects of the
initial state radiation and only consider the final state collinear
divergences.

We first write the phase space integral in the usual form
\begin{eqnarray}
        \hspace*{-10pt}
        \int\!dP\!S & = & \frac{1}{(2\pi)^8}
                \int\frac{d^3 q_-}{2q_{-}^{0}}
                \int\frac{d^3 q_+}{2q_{+}^{0}}
                \int\frac{d^3 k_1}{2k_{1}^{0}}
                \int\frac{d^3 k_2}{2k_{2}^{0}}
                \,\delta^{(4)}(p_- + p_+ - q_- - q_+ - k_1 - k_2)
\nonumber\\
                        & = &
                \frac{1}{2\pi}\int\!ds_{-1}
                \frac{1}{2\pi}\int\!ds_{+2}
                \int\!d\Omega \frac{\sqrt{\lambda(s,s_{-1},s_{+2})}}{32\pi^2 s}
                \int\!d\Omega_1 \frac{s_{-1} - m_\mu^2}{32\pi^2 s_{-1}}
                \int\!d\Omega_2 \frac{s_{+2} - m_\mu^2}{32\pi^2 s_{+2}}
\nonumber\\
\end{eqnarray}
with $s_{\pm i} = (q_\pm+k_i)^2 = m_\mu^2 + 2q_\pm\inp k_i$ and $d\Omega_i =
d\phi_i\,d\cos\theta_i$, all angles being defined in their respective center
of mass systems; $\lambda(s,s_{-1},s_{+2}) = (s - s_{-1} - s_{+2})^2 -
4s_{-1}s_{+2}$ is the Kall\`en function of the first two-particle phase space.
Actually, the first five integrations, up to $d\phi_1$, are already in a
useful form.  The last three will have to be rewritten.  In a CM frame in
which $({\vec{q}}_-+{\vec{k}}_1)$ is parallel to the $z$-axis and
$q_-^y = 0$ we find for
the remaining invariants, including $s_{12} = (k_1+k_2)^2$:
\begin{eqnarray}
        s_{+1} 
                        & = & m_\mu^2 - s_{12} +
                \frac{1}{2} \bigl(1-\frac{m_\mu^2}{s_{-1}}\bigr)
                \bigl(s-s_{-1}-s_{+2} - \sqrt{\lambda} \cos\theta_1 \bigr)
\\
\label{eq:s-2}
        s_{-2} 
                        & = & m_\mu^2 - s_{12} +
                \frac{1}{2} \bigl(1-\frac{m_\mu^2}{s_{+2}}\bigr)
                \bigl(s-s_{-1}-s_{+2} + \sqrt{\lambda} \cos\theta_2 \bigr)
\\
        s_{12} 
                        & = & \frac{1}{4}
                \bigl(1-\frac{m_\mu^2}{s_{-1}}\bigr)
                \bigl(1-\frac{m_\mu^2}{s_{+2}}\bigr)
                \bigl( (s-s_{-1}-s_{+2})(1-\cos\theta_1\cos\theta_2) -
                        \sqrt{\lambda}(\cos\theta_1 - \cos\theta_2)
\nonumber\\&&\mbox{}
                - 2\sqrt{s_{-1}s_{+2}}\sin\theta_1\sin\theta_2\cos\phi_2 \bigr)
\label{eq:s12}
\end{eqnarray}
We thus have to transform the $\phi_2$ integral into an integral over $s_{12}$
and change the order of integration of this integral with the two
$\cos\theta_i$ integrals, which can then be replaced by integrals over
$s_{+1}$ and $s_{-2}$, respectively.  Let's start with the first exchange:
\begin{eqnarray}
        \lefteqn{\int_{-1}^{+1}\!d\,\cos\theta_2 \int_0^{2\pi}\!d\,\phi_2 =
                \int_{-1}^{+1}\!d\,\cos\theta_2 \int_{-1}^{+1}\!d\,\cos\phi_2
                        \Bigl(\frac{1}{\sin\phi_2}\Bigr|_{\sin\phi_2>0} -
                             \frac{1}{\sin\phi_2}\Bigr|_{\sin\phi_2<0} \Bigr)}
\\
        & = & \int_{-1}^{+1}\!d\,\cos\theta_2
        \int_{s_{12}^-(\cos\theta_1,\cos\theta_2)}^{s_{12}^+(\cos\theta_1,
                \cos\theta_2
 }
        \!\!\!d\,s_{12} \frac{2}{(1-m_\mu^2/s_{-1})(1-m_\mu^2/s_{+2})
                \sqrt{s_{-1}s_{+2}}
                \sin\theta_1\sin\theta_2}\,\frac{2}{\sin\phi_2}
\nonumber
\end{eqnarray}
In the last step we assume that the function to be integrated is symmetric
under the exchange $\phi_2\to-\phi_2$, if not, one will have to keep the
orginal form.  The matrix element of the previous section does not have this
property, but for simplicity we keep the simplified notation.  The functions
$s_{12}^\pm(\cos\theta_1,\cos\theta_2)$ are just given by eq.\ (\ref{eq:s12})
for $\cos\phi_2 = \mp1$.  They assume their extrema for $\cos\theta_2 = \mp
\bigl((s-s_{-1}-s_{+2})\cos\theta_1 - \sqrt{\lambda}\bigr)/
\bigl((s-s_{-1}-s_{+2}) - \sqrt{\lambda}\cos\theta_1\bigr)$,
giving $s_{12}^-(\cos\theta_1) = 0$ and $s_{12}^+(\cos\theta_1) =
(1-m_\mu^2/s_{-1})(1-m_\mu^2/s_{+2})(s-s_{-1}-s_{+2}
-\sqrt{\lambda}\cos\theta_1)/2$.  One thus obtains
\begin{eqnarray}
        & = & \int_0^{s_{12}^+(\cos\theta_1)} \!\!d\,s_{12}
                \int_{\cos\theta_2^-}^{\cos\theta_2^+} \!\!d\,\cos\theta_2
                \frac{2}{(1-m_\mu^2/s_{-1})(1-m_\mu^2/s_{+2})
                \sqrt{s_{-1}s_{+2}}
                \sin\theta_1\sin\theta_2}\,\frac{2}{\sin\phi_2}
\nonumber\\
\end{eqnarray}
where the bounds $\cos\theta_2^\pm(s_{12})$ follow from the inversion of
$s_{12}^\pm(\cos\theta_1,\cos\theta_2)$.  This integration region is shown for
two values of $\cos\theta_1$ in fig.\ \ref{fig:phasespace}.  The integration
of $\cos\theta_2$ is now trivially transformed into an integration over
$s_{-2}$ using eq.\ (\ref{eq:s-2}).

\begin{figure}[htb]
\centerline{
\unitlength .5bp
\setepsfscale{.5}
\begin{picture}(770,554)(0,0)
\put(0,0){\strut\epsffile{phasespace.ps}}
\put(23.4,468.9){\makebox(0,0)[tr]{\shortstack{\large$s_{12}$\\\large$
    [\GEV]$}}}
\put(696.9,25.1){\makebox(0,0)[tr]{\large$\cos\theta_2$}}
\end{picture}
\unitlength 1bp
\setepsfscale{1}
}
\caption{Phase space in $s_{12}$ and $\cos\theta_2$ for some value of
$s_{-1},s_{+2}$ and two values of $\cos\theta_1$.}
\label{fig:phasespace}
\end{figure}

Swapping the order of $\cos\theta_1$ and $s_{12}$
integrals is done analogously:
\begin{equation}
        \int_{-1}^{+1}\!d\,\cos\theta_2 \int_0^{s_{12}^+(\cos\theta_1)}
                \!\!d\,s_{12}
         = \int_0^{s_{12}^+} \!\!d\,s_{12}
                \int_{-1}^{\cos\theta_1^+} d\,\cos\theta_1
\end{equation}
with
\begin{eqnarray}
        s_{12}^+ & = & \frac{1}{2}\bigl(1-\frac{m_\mu^2}{s_{-1}}\bigr)
                \bigl(1-\frac{m_\mu^2}{s_{+2}}\bigr)\, \bigl(s-s_{-1}-s_{+2}+
                \sqrt{\lambda}\bigr)\\
        \cos\theta_1^+ & = & \min\biggl(1,\frac{s-s_{-1}-s_{+2}-
            \frac{2s_{12}}{(1-m_\mu^2/s_{-1})(1-m_\mu^2/s_{+2})}}
       {\sqrt{\lambda}}\biggr)
\end{eqnarray}
Again, a simple rescale now suffices to obtain the required integral over
$s_{+1}$.

In the course of these transformations great care must be taken to guard
against numerical instabilities in the limit $s_{\pm i} \ll s$, which is the
collinear region.  In effect, one can lose a factor $m_\mu^8/s^4 \approx
3\cdot10^{-24}$ in the computation of the bounds on $\cos\theta_2$.


\section{Results}

We have integrated the lowest order and initial state corrected matrix
elements of sect. \ref{sec:ME} using the kinematics described in sect.\
\ref{sec:kin}.  For the initial state bremsstrahlung we used the leading
log expression of ref.\ \ci{[2]}, which takes into account up to two
hard collinear photons and $n$ soft ones.  To convert to absolute
numbers we assumed that all the 800000 observed $Z$'s (corresponding
to 950000 produced) which were used in the analysis were produced on the
peak.  This corresponds to an effective integrated luminosity of $13.29
\pB^{-1}$ when initial state radiation is not taken into account and
$18.15 \pB^{-1}$ when it is.  Furthermore, we used the same cuts as
described in \ci{[1]}:
\begin{eqnarray}
\label{eq:cutEgam}
        \tilde{E}_\gamma  & > & 1 \GeV \\
        E_\mu     & > & 3 \GeV \\
        |\cos\theta_\gamma| & < & 0.9 \\
        |\cos\theta_\mu| & < & \cos 36^\circ \\
        \cos\theta_{\mu\gamma} & < & \cos 5^\circ
\label{eq:cutcosmg}
\end{eqnarray}
with $\tilde{E}_\gamma$ the energy of the $\gamma$ smeared with a 2\%
gaussian.  In spite of the $E_\gamma^{-4}$ behaviour of the matrix element the
effect of this smearing is negligible.  We did not implement any cut on
the $\gamma\gamma$ separation as the criteria for this were not easily
tanslated into a simple cut.  However, as there is no divergence associated
with this angle the effect of neglecting it will be minimal.

We find that, correcting for the $Z$ peak height, the effects of
collinear initial
state radiation are negligible for all observables we consider except the
cross section for low-mass muon pairs.  In the following we will only give the
results including initial state bremsstrahlung.  Furthermore, even for the
relatively small $\mu\gamma$ cut-off angle used here there are no discernable
mass effects.  Because the cut on $E_\gamma$ is comparable to the width
of the $Z$ peak, the initial state \{I\} and mixed \{IF\} diagrams
contribute strongly to the total cross section (15\%).  However,
as expected, they are a
factor $20 \approx 2\Gamma_{Z}/M_Z$ suppressed with respect to the final
state radiation in the high $\mg$ region ($\mg>20\GeV$). Monte
Carlo programs only taking into account final state radiation (like the one
described in \ci{[4]}) are thus expected to give a reasonable
description for the high mass events.

The $Z$ parameters we used for the calculation are $\alpha(M_Z) =
1/128.5$, $M_Z =91.174 \GeV$, $\Gamma_{Z} = 2.50 \GeV$ and $\Gamma_e =
0.08324 \GeV$ \cite{[11]}.  For the coupling of the two extra photons we
took $\alpha_0 = 1/137$.  The total cross section of the reaction
$e^+e^- \to \mu^+\mu^-\gamma\gamma$ after the cuts
(\ref{eq:cutEgam}-\ref{eq:cutcosmg}) is $4.43 \pB$.  This translates
into 80 expected events.  This is somewhat more than the 68 observed and
56 predicted quoted in ref.\ \ci{[1]}.

Comparing $d\sigma/d\mg$ for muons only, fig.\ \ref{fig:mgg}, with the
Monte Carlo histogram of ref.\ \ci{[1]}, which sums muons, electrons and
taus, there are large differences.  The low mass region cannot be
compared because of different cuts on the three channels.  However, in
the high mass region ($\mg > 40 \GeV$) we obtain for muons only already
75\% of the number given in \cite{[1]} for all three channels together.
This difference is most likely
due to the leading log approximation used in the YFS3 Monte Carlo
\cite{[5]}, which is appropriate for the large majority of events, but
not for this region.  When we compare the expected and observed number
of events for several lower cuts on $\mg$ we obtain tab. \ref{table}.
We conclude that the {\it number\/} of $\mu^+\mu^-$ events is entirely
explainable within QED.

\begin{table}[htb]
\centerline{
\begin{tabular}{|r|l|r@{.}l|r|r@{.}l|}
\hline
$\mg^{\rm min}$
    & \multicolumn{1}{|c|}{$\sigma$}
             & \multicolumn{2}{|c|}{$n_{\rm exp}$}
                     & $n_{\rm obs}$
                          & \multicolumn{2}{|c|}{$P(n\geq n_{\rm obs})$} \\
\multicolumn{1}{|c|}{$[\GEV]$}
    & \multicolumn{1}{|c|}{$[\PB]$}
             & \multicolumn{2}{|c|}{}
                     &    & \multicolumn{2}{|c|}{\%}  \\
\hline
 0  & 4.43   &  80&  & 68 & \multicolumn{2}{|c|}{}  \\
40  & 0.0816 &  1&48 &  3 & ~~~~~18& \\
50  & 0.0351 &  0&64 &  3 & 2&7 \\
55  & 0.0225 &  0&41 &  3 & 0&8 \\
60  & 0.0141 &  0&26 &  1 & 23& \\
65  & 0.0080 &  0&14 &  0 & 86& \\
\hline
\end{tabular}
}
\caption{Comparison of the expected and observed number of events for different
$\mg^{\rm min}$.}
\label{table}
\end{table}

\begin{figure}[t]
\centerline{
\begin{picture}(554,504)(0,0)
\put(0,0){\strut\epsffile{fig6.ps}}
\put( 70.8,468.9){\makebox(0,0)[tr]{\shortstack{\large\strut
events\\\large\strut per\\\large\strut$2.5\GeV$}}}
\put(480.8, 19.1){\makebox(0,0)[tr]{\large$\mg[\GEV]$}}
\end{picture}
}
\caption[]{Expected number of events per 2.5 GeV bin in $\mg$, assuming an
integrated
luminosity of $18.15 \pB^{-1}$ at $\sqrt{s} = 91.2 \GeV$.}
\label{fig:mgg}
\end{figure}

To comment on the probability of observing the 4 events clustered at one
particular mass value, we have to include the $e^+e^-$ and $\tau^+\tau^-$
final states.  However, the first one necessitates the inclusion of t-channel
graphs whereas the latter requires a thorough understanding of the detector.
We have not performed these studies, hence the only conclusion we can draw is
that the probablility will certainly be higher than the one reported in
\ci{[1]}.

In figs \ref{fig:mllmgg} and \ref{fig:costmgg} we give the distributions
which are analogous to figs 4 and 5 in ref.\ \ci{[1]}
\footnote{Notice that two of the four high $\mg$ events are rather collinear
(at small $\theta_{\mu\gamma}$) and that the YFS3 Monte Carlo simulation
just produced one collinear hard $\gamma \gamma$ event.
}.  The only
feature which differs from a smooth exponential behaviour is the ridge for low
invariant muon masses, this is due to the process $e^+e^- \to
\gamma\gamma\gamma^*, \gamma^* \to \mu^+\mu^-$, which we unknowingly included.
The total cross section for events with an invariant muon pair mass
$m_{\mu\mu} < 5 \GeV$ is 0.04 pb, or 0.7 expected events.  As can be
seen from fig.\ \ref{fig:mllmgg} these are roughly evenly ditributed
between $\mg = 40$ and 80 GeV, with a total of 4 fb in this region.  One
thus has a one in four chance of observing such an event at one of the
LEP detectors.

\begin{figure}[p]
\setepsfscale{.5}
\unitlength .5bp
\centerline{
\begin{picture}(554,384)(0,0)
\put(0,0){\strut\epsffile{fig4.ini.ps}}
\put( 70.8,468.9){\makebox(0,0)[tr]{\shortstack{\large\strut
$\mg$\\\large\strut$[\GEV]$}}}
\put(480.8, 19.1){\makebox(0,0)[tr]{\large$m_{\mu\mu}[\GEV]$}}
\end{picture}
}
\centerline{
\setepsfscale{.7}
\unitlength .7bp
\begin{picture}(554,474)(-50,0)
\put(0,0){\strut\epsffile{fig4.ini.3d.hacked.ps}}
\put( 70.8,468.9){\makebox(0,0)[tr]{\shortstack{\Large\strut
$\frac{d^2\sigma}{d\mg dm_{\mu\mu}}$\\\large\strut$[\PB/\GEV^2]$}}}
\put( 70.8, 110){\makebox(0,0){\large$m_{\mu\mu}[\GEV]$}}
\put(480.8, 110){\makebox(0,0)[tr]{\large$\mg[\GEV]$}}
\end{picture}
}
\caption{Scatterplot and differential cross section for the photon-photon
invariant mass
versus the muon-muon invariant mass.  Each dot corresponds to
0.2 event.}
\label{fig:mllmgg}
\end{figure}

\begin{figure}[p]
\setepsfscale{.5}
\unitlength .5bp
\centerline{
\begin{picture}(554,384)(0,0)
\put(0,0){\strut\epsffile{fig5.ini.ps}}
\put( 70.8,468.9){\makebox(0,0)[tr]{\large$\cos\theta_{\mu\gamma}^{\rm min}$}}
\put(480.8, 19.1){\makebox(0,0)[tr]{\large$\mg[\GEV]$}}
\end{picture}
}
\centerline{
\setepsfscale{.7}
\unitlength .7bp
\begin{picture}(554,474)(-50,0)
\put(0,0){\strut\epsffile{fig5.ini.3d.hacked.ps}}
\put( 70.8,468.9){\makebox(0,0)[tr]{\shortstack{\Large\strut
$\frac{d^2\sigma}{d\cos\theta d\mg}$\\\large\strut$[\PB/\GEV]$}}}
\put( 40.8, 140){\makebox(0,0){\large$\cos\theta_{\mu\gamma}^{\rm min}$}}
\put(480.8,  70){\makebox(0,0)[tr]{\large$\mg[\GEV]$}}
\end{picture}
}
\caption{Scatterplot and differential cross section for the cosine of the
smallest angle
between a muon and a photon versus the photon-photon invariant
mass.
Each dot corresponds to 0.2 event.}
\label{fig:costmgg}
\end{figure}


\section{Conclusions}

We find a significantly higher cross section for high mass
$\mu^+\mu^-\gamma\gamma$ events at the $Z$ peak than is given as the
theoretical prediction in ref. \ci{[1]}.  The difference is probably due to
the leading log approximation used in the program employed to generate this
curve.
As we have not studied the final states with electrons and taus we cannot
compute the chances of the four events clustering around one mass value.
However, the other collaborations do
not see any clustering, and also considering the number of plots considered
which do not give rise to anything striking, it seems compatible
with QED. There is therefore no urgent need to search for models beyond the
standard model to explain this behaviour.  Higher statistics will provide the
experimental test of this explanation.

When finishing this work we obtained a Report \ci{[12]} in which several
generators are compared. Our results are consistent with the numbers at the
upper end of the range reported.

{\em Acknowledgment. }

One of us (KK) would like to thank the Paul Scherrer Institute
for the kind hospitality during his stay at Villigen where this work was
started.


\bb{19}

\bi{[1]} The L3 Collaboration (O. Adriani et al.), Phys. Lett. B295 (1992)
         337.

\bi{[2]}
F. A. Berends, G. Burgers, W. L. van Neerven, Phys. Lett. B177 (1986) 191;
Nucl. Phys. B297 (1988) 429; E: Nucl. Phys. B304 (1988) 921; \\
W. Beenakker, F. A. Berends, W.L. van Neerven, in: Proc. Workshop
on Electroweak Radiative Corrections (Ringberg Castle, FRG), ed.
J. H. K\"uhn, Springer Berlin 1989; F. A. Berends et al., in
{\em Z Physics at LEP1}, eds. G. Altarelli et al., CERN 89-08 (1989).

\bi{[3]} S. Jadach, B. F. L Ward, E. Richter-Was, H. Zhang,
         Phys. Rev. D 42 (1990) 2977.

\bi{[4]} W. J. Stirling, Phys. Lett. B271 (1991) 261;
         Z. Kunszt, unpublished

\bi{[5]} S. Jadach, B. F. L Ward, Phys. Lett. B274 (1992) 470.

\bi{[6]}
K. Ko\l odziej, M. Zra\l ek, Phys. Rev. D43 (1991) 3619.

\bi{[7]}
F. A. Berends, W. T. Giele, Nucl. Phys. B306 (1988) 759; B313 (1989) 595; \\
F. A. Berends, W. T. Giele and H. Kuijf, Phys. Lett.
B232 (1989) 266; Nucl. Phys. B321 (1989) 39; B333 (1990) 120.

\bibitem{Jacob&Wick} M. Jacob and G.C. Wick,\newblock Ann. Phys. (N.Y.) {
B82} (1959) 404.

\bibitem{Form} J. Vermasseren, {\it FORM}, Computer Algebra Nederland,
Amsterdam, 1992.

\bi{[10]}
G. P. Lepage, J. Comp. Phys. 27 (1978) 192; Cornell
University preprint, CLNS-80/447 (1980).

\bi{[11]} The LEP Collaborations: ALEPH, DELPHI, L3 and OPAL, Phys. Lett.
B276 (1992) 247.

\bi{[12]} K. Riles, Report UM-HE-92-36, L3
\underline{Internal} Note \#1293.

\end{thebibliography}

\end{document}